\documentclass[prd,twocolumn,floatfix,amsmath,nofootinbib,amssymb,floatfix]{revtex4}
\usepackage{graphicx,color,dcolumn,booktabs,bm}
\usepackage{longtable,lscape}
\usepackage{txfonts}
\usepackage{overpic}
\usepackage{amssymb}
\usepackage{indentfirst}
\usepackage{feynmf}   
\usepackage{slashed}  
\usepackage{cases}
\usepackage{color}
\usepackage{multirow}
\usepackage{epstopdf}
\usepackage{enumerate}
\usepackage[colorlinks, citecolor=blue,anchorcolor=red,menucolor=red, linkcolor=red,filecolor=red,urlcolor=blue,frenchlinks=red]{hyperref}

\begin{document}

\title{Universal non-resonant explanation to charmoniumlike structures $Z_c(3885)$ and $Z_c(4025)$}
\author{Jun-Zhang Wang$^{1,2}$}\email{wangjzh2012@lzu.edu.cn}
\author{Dian-Yong Chen$^3$}\email{chendy@seu.edu.cn}
\author{Xiang Liu$^{1,2}$\footnote{Corresponding author}}\email{xiangliu@lzu.edu.cn}
\author{Takayuki Matsuki$^{4}$}\email{matsuki@tokyo-kasei.ac.jp}
\affiliation{$^1$School of Physical Science and Technology, Lanzhou University, Lanzhou 730000, China\\
$^2$Research Center for Hadron and CSR Physics, Lanzhou University $\&$ Institute of Modern Physics of CAS, Lanzhou 730000, China\\
$^3$School of Physics, Southeast University, Nanjing 210094, China\\
$^4$Tokyo Kasei University, 1-18-1 Kaga, Itabashi, Tokyo 173-8602, Japan
}

\date{\today}

\begin{abstract}
Different from the usual tetraquark assignment to charged $Z_c(3885)$ and $Z_c(4025)$ charmoniumlike structures, 
in this letter we propose a universal non-resonant explanation to decode these $Z_c$'s,
which is based on a special 
dynamical behavior of $e^+e^-\to D^{(*)}\bar{D}^*\pi$. 
Our study shows that $Z_c(3885)$ and $Z_c(4025)$ are only the reflection from the $P$-wave charmed meson $D_1(2420)$ involved in $e^+e^-\to D^{(*)}\bar{D}^*\pi$. 
Obviously, the present work provides a unique perspective, which 
can be examined by future experiments like BESIII and BelleII. 
\end{abstract}

\maketitle

\noindent\textit{Introduction.}--Exploring exotic hadrons is a hot issue in hadron physics. Especially 
 with more and more observations of charmoniumlike $XYZ$ states, theorists have carried out extensive study since 2003, which really
 deepens on our understanding of how these novel phenomena happen (see review articles \cite{Chen:2016spr,Chen:2016qju,Liu:2019zoy,Guo:2017jvc,Olsen:2017bmm,Brambilla:2019esw} for learning the relevant progress). Of course, our knowledge of non-perturbative quantum chromodynamics 
 (QCD) becomes more abundant. 
 
Among these investigations, it is a key point how to identify exotic hadrons as definite quark states from the observed charmoniumlike $XYZ$ states, which is not an easy task. 
 Before establishing the existence of exotic states, we need to exhaust the possibilities of explaining them in the conventional framework. Here, we need to check whether  
the observed charmoniumlike $XYZ$ states can be categorized into a conventional hadron family. A typical example is $X(3915)$ and $Z(3930)$ produced from the photon-photon fusion process \cite{Uehara:2009tx,Uehara:2005qd}, which can be assigned to charmonia $\chi_{c0}(2P)$ and $\chi_{c2}(2P)$ \cite{Liu:2009fe}, respectively. In experimental analysis, experiment usually claims the observation of some state if a resonance structure exists in the corresponding invariant mass spectrum. After the observation of two charomiumlike structures $Y(4260)$ and $Y(4360)$, the Lanzhou group indicated that the line shapes of $Y(4260)$ and $Y(4360)$ 
can be reproduced by the interference effect of two charmonia, $\psi(4160)$ and $\psi(4415)$, and the continuum contribution \cite{Chen:2015bft}, by which the puzzling phenomenon of the missing of $Y(4260)$ and $Y(4360)$  in the open-charm decay channels and the $R$ value measurement  can be understood. This study reflects that {\it "what you see is not what you get"}. 

Along this line, in this work we propose a universal non-resonant explanation to charged charmoniumlike structures $Z_c(3885)$ and $Z_c(4025)$. Here, the charged $Z_c(3885)$ and $Z_c(4025)$ were reported by the BESIII Collaboration when analyzing the three-body open-charm decay channels from $e^+e^-$ annihilation, $e^+e^-\to \pi^\pm(D\bar{D^*})^\mp$ \cite{Ablikim:2013xfr} and $e^+e^-\to \pi^\pm(D^*\bar{D^*})^\mp$ \cite{Ablikim:2013emm}, respectively.  Additionally, the neutral partners of $Z_c(3885)$ and $Z_c(4025)$ were also found in the $e^+e^-\to \pi^0(D\bar{D^*})^0$ \cite{Ablikim:2015gda} and $e^+e^-\to \pi^0(D^*\bar{D^*})^0$ \cite{Ablikim:2015vvn} processes. Since $Z_c(3885)$ and $Z_c(4025)$ are charged states, it is easy to conjugate that $Z_c(3885)$ and $Z_c(4025)$ should contain at least four quarks, which can be good candidates of exotic tetraquark states \cite{He:2013nwa,Braaten:2014qka,Maiani:2014aja,Guo:2013sya,Karliner:2015ina,Voloshin:2013dpa,Aceti:2014uea,Wang:2013vex,Deng:2014gqa,Chen:2015ata,Goerke:2016hxf,Patel:2014vua,Wang:2013exa,Qiao:2013dda,Agaev:2020zad,Agaev:2017tzv,Agaev:2016dev}, which has become a popular and dominant opinion since 2014. {In addition, the kinematical explanations of $Z_c(3885)$ as an enhancement from triangle singularity have also been widely discussed in Refs. \cite{Cleven:2013mka,Qin:2016spb,Liu:2015taa,Wang:2013cya,Guo:2019twa}. }

Totally different from this tetraquark state assignment to $Z_c(3885)$ and $Z_c(4025)$, 
in this work, we indicate that $Z_c(3885)$ and $Z_c(4025)$ are only the reflection of the $P$-wave charmed meson $D_1(2420)$ involved in $e^+e^-\to D^{(*)}\bar{D}^*\pi$, which is due to 
a special dynamical behavior of three-body process from $e^+e^-$ annihilation. The details will be illustrated in the following sections.  
By this realistic and novel example, we want to show that what you see is not always what you get. It is obvious that the present work provides a unique perspective
to decode charged charmoniumlike structures $Z_c(3885)$ and $Z_c(4025)$, which should be emphasized before definitely identifying them as exotic states.

\noindent\textit{A general analysis of $e^+e^-\to D^{(*)}\bar D_1(2420)\to D^{(*)}\bar D^*\pi$.}--In the electron and positron annihilation process, the two-body open charm channels $D^{(*)}\bar{D}^{**}$ due to a direct coupling with vector charmonium or charmoniumlike states usually play an important role in producing three-body open charm final states $D^{(*)}\bar{D}^{(*)}M$, where $\bar{D}^{**}$ and $M$ stand for a higher excited charmed meson and a light meson, respectively. Hence, this production mechanism is also expected to exist in the process $e^+e^- \to (D\bar{D}^*)^{\pm}\pi^{\mp}$ and $e^+e^- \to (D^*\bar{D}^*)^{\pm}\pi^{\mp}$, where the intermediate resonance $\bar{D}^{**}$  can be directly expressed by a Breit-Wigner distribution in the invariant mass spectrum of $\bar{D}^{(*)}\pi$. However, the dynamical and kinematical behaviors owing to some special higher excited charmed mesons may cause a corresponding line shape of a reflective peak on a differential cross section vs. the invariant mass of $D^{(*)}\bar{D}^{(*)}$, which can provide a  new perspective to revisit the nature of two charmoniumlike states $Z_c(3885)$ and $Z_c(4025)$ in the invariant mass distribution of $D^{(*)}\bar{D}^{(*)}$.

In the charmed meson family, $D_1(2420)$ with $J^{P}=1^{+}$ is a relatively narrow resonance, whose main reason is that $D_1(2420)$ can only decay into a dominant final state $D^*\pi$ via a $D$-wave due to the limit of heavy quark symmetry.  We have noticed that the averaged resonance widths of $D_1(2420)$ listed in Particle Data Group (PDG) are $25\pm6$ MeV for charged state and $31.7\pm2.5$ MeV for neutral state  \cite{Tanabashi:2018oca}, which are very close to experimental widths of $24.8\pm3.3\pm11.0$ MeV of $Z_c(3885)^{\pm}$ \cite{Ablikim:2013xfr} and $24.8\pm5.6\pm7.7$ MeV of $Z_c(4025)^{\pm}$ \cite{Ablikim:2013emm}. In addition, the processes $e^+e^- \to (D\bar{D}^*)^{\pm}\pi^{\mp}$ and $e^+e^- \to (D^*\bar{D}^*)^{\pm}\pi^{\mp}$ are measured at center of mass energy of 4.26 GeV by BESIII \cite{Ablikim:2013xfr,Ablikim:2013emm}, which is below the thresholds of both channels $D\bar D_1(2420)$ and $D^{*}\bar D_1(2420)$. 
Since the measuring energy, 4.26 GeV, is close to the threshold of $D^{(*)}\bar D^*$, we can expect to see a reflection phenomenon of $D_1(2420)$ with peaks.
The above properties indicate that the $P$-wave charmed meson $D_1(2420)$ is an excellent candidate for providing a possible universal non-resonant explanation to $Z_c(3885)$ and $Z_c(4025)$. Therefore, in the following, we will perform a general analysis for the process  $e^+e^-\to D^{(*)}\bar D_1(2420)\to D^{(*)}\bar D^*\pi$ based on the framework of an effective Lagrangian approach.

\begin{figure}[htbp]
	\includegraphics[width=8.6cm,keepaspectratio]{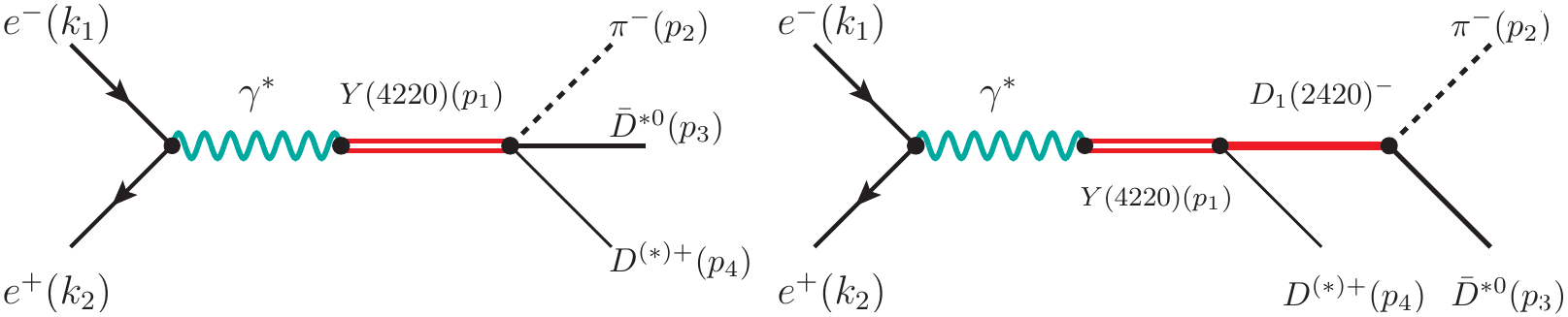}
	\caption{  The schematic diagrams of two main production contributions in the three-body open charm processes $e^+e^- \to D^{(*)+}\bar D^{*0}\pi^-$.
	\label{fig:feydiagram} }
\end{figure}

     The schematic diagrams of the processes, $e^+(k_2)e^-(k_1)\to Y(4220)(p_1) \to D^+(p_4)\bar D_1(2420)^-\to D^+(p_4)\bar D^{*0}(p_3)\pi^-(p_2)$ and $e^+(k_2)e^-(k_1)\to Y(4220)(p_1) \to D^{*+}(p_4)\bar D_1(2420)^-\to D^{*+}(p_4)\bar D^{*0}(p_3)\pi^-(p_2)$, are shown in Fig. \ref{fig:feydiagram}, where $p_i$'s in brackets represent the corresponding four-momentum of particles in the rest frame of $e^+e^-$ system. Additionally, the background diagrams of $e^+e^-\to Y(4220) \to  D^+\bar D^{*0}\pi^-$ and $e^+e^-\to Y(4220) \to D^{*+}\bar D^{*0}\pi^-$ are also considered. {Here, there exists a direct coupling between virtual photon generated by the electron and positron annihilation and an observed charmoniumlike structure $Y(4220)$ \cite{Ablikim:2016qzw} near 4.26 GeV, which has been recently confirmed in the measurement of a total cross section of $e^+e^- \to D^0D^{*-}\pi^+$ by BESIII \cite{Ablikim:2018vxx}. In our previous studies to the $Y$ problem \cite{Wang:2019mhs}, the nature of $Y(4220)$ can be well understood by a lower mass charmonium state from the mixture of components $\psi(4S)$ and $\psi(3D)$ in an unquenched potential model. Then, it is very natural that the $Y(4220)$ as a charmonium state has a strong coupling with two-body open charm decay channels $D_1(2420)D^{(*)}$ via an $S$-wave interaction. On the other hand, it is worth mentioning that when explaining $Y(4220)$ as a configuration of a $DD_1(2420)$ molecular state \cite{Cleven:2013mka,Qin:2016spb,Wang:2013cya}, the contribution of an intermediate $D_1(2420)^-D^{*+}$ channel from the coupling with $Y(4220)$ in Fig. \ref{fig:feydiagram} may not be dominant for the reaction $e^+e^- \to D^{*+}\bar D^{*0}\pi^-$.   }

All the related Lagrangian densities for calculating the reaction processes in Fig. \ref{fig:feydiagram}  are listed below \cite{Bauer:1975bv,Bauer:1975bw,Casalbuoni:1996pg,Chen:2011xk,Liu:2020ruo} and
\begin{eqnarray}
\mathcal{L}_{\gamma Y}&=&\frac{-em_Y^2}{f_Y}Y_{\mu}A^{\mu}, \\
\mathcal{L}_{YD{D}^*\pi}&=&g_{YD{D}^*\pi}Y_{\mu}(D^{\dag}\boldsymbol{\tau} \cdot \boldsymbol{\pi} D^{*\mu}+D^{*\mu\dag}\boldsymbol{\tau} \cdot \boldsymbol{\pi} D), 
\end{eqnarray}
\begin{eqnarray}
\mathcal{L}_{YD^*{D}^*\pi}&=&-ig_{YD^*{D}^*\pi}\varepsilon^{\mu\nu\alpha\beta} (Y_{\mu}D^{*\dag}_{\nu}\partial_{\alpha}\boldsymbol{\tau} \cdot \boldsymbol{\pi} D^{*}_{\beta}   \nonumber \\
&&+\partial_{\mu}Y_{\nu}D^{*\dag}_{\alpha}\boldsymbol{\tau} \cdot \boldsymbol{\pi} D^{*}_{\beta}),  \\
\mathcal{L}_{D_{1}{D}Y}&=&g_{D_{1}{D}Y}(D_{1\mu}^{\dag}D-D^{\dag}D_{1\mu})Y^{\mu},    \\
\mathcal{L}_{D_{1}{D}^*Y}&=&ig_{D_{1}{D}^*Y}\varepsilon^{\mu\nu\alpha\beta}(D_{1\mu}^{\dag}D^*_{\nu}-D^{*\dag}_{\nu}D_{1\mu})\partial_{\alpha}Y_{\beta},   \\
\mathcal{L}_{{D}_{1}D^*\pi}&=&g_{{D}_{1}D^*\pi}(3D_{1}^{\mu\dag}(\partial_{\mu}\partial_{\nu}\boldsymbol{\tau} \cdot \boldsymbol{\pi})D^{*\nu}-D_{1}^{\mu\dag}(\partial_{\nu}\partial_{\nu}\boldsymbol{\tau} \cdot \boldsymbol{\pi})D^{*\mu} \nonumber \\
&&+3D^{*\nu\dag}(\partial_{\mu}\partial_{\nu}\boldsymbol{\tau} \cdot \boldsymbol{\pi})D_{1}^{\mu}-D^{*\mu\dag}(\partial_{\nu}\partial_{\nu}\boldsymbol{\tau} \cdot \boldsymbol{\pi})D_{1}^{\mu}),
\end{eqnarray}
where $A^{\mu}$, $Y$, $D_{1}^{\mu}$, $\boldsymbol{\pi}$ are photon, charmoniumlike state $Y(4220)$, charmed meson $D_{1}(2420)$, and pion fields, respectively, and $g_{YD{D}^*\pi}$, $g_{YD^*{D}^*\pi}$, $g_{D_{1}{D}Y}$, $g_{D_{1}{D}^*Y}$, and $g_{{D}_{1}D^*\pi}$ are the coupling constants, and $\boldsymbol{\tau}$ represents the Pauli matrix. Though some of coupling constants in the above Lagrangian densities are unknown, they do not affect our predictions for the line shapes of differential cross sections.
Based on the above interaction vertices, the general amplitudes of four processes in Fig. \ref{fig:feydiagram}  can be written as
\begin{eqnarray}
\mathcal{M}_{(a)}^{Nonpeak}&=&\mathcal{A}^{e^+e^- \to Y(4220)}_{\rho} g_{YD{D}^*\pi}\epsilon_{D^*}^{\rho *},  \label{np-a}  \\
\mathcal{M}_{(a)}^{D_1(2420)}&=&\mathcal{A}^{e^+e^- \to Y(4220)}_{\rho}g_{D_{1}{D}Y} g_{{D}_{1}D^*\pi}\epsilon_{D^*}^{\lambda *} \nonumber  \\
&&\times \frac{\tilde{g}^{\rho}_{\alpha}(3p_2^{\alpha}p_{2\lambda}-g^{\alpha}_{ \lambda}p_2^{\tau}p_{2\tau})}{(p_1-p_4)^2-m_{D_{1}}^2+im_{D_{1}}\Gamma_{D_{1}}}, \label{d1-a}  \\
\mathcal{M}_{(b)}^{Nonpeak}&=&\mathcal{A}^{e^+e^- \to Y(4220)}_{\rho} (-ig_{YD^*{D}^*\pi})\varepsilon^{\omega\lambda\alpha\beta}\epsilon_{D^*}^{\delta *} \nonumber \\
&&\times (ip_{2\alpha}g_{\delta\lambda}g_{\omega}^{\rho}-ip_{1\omega}g_{\delta\alpha}g_{\lambda}^{\rho})\epsilon_{D^*\beta}^{ *},  \label{np-b}  \\
\mathcal{M}_{(b)}^{D_1(2420)}&=&\mathcal{A}^{e^+e^- \to Y(4220)}_{\rho}ig_{D_{1}{D}^*Y}g_{{D}_{1}D^*\pi}\varepsilon^{\omega\alpha\beta\rho} \epsilon_{D^*\alpha}^{ *} \nonumber \\
&&\times (-ip_{1\beta}) \frac{\tilde{g}_{\omega\theta}(3p_2^{\theta}p_{2\lambda}-g^{\theta}_{ \lambda}p_2^{\tau}p_{2\tau}) \epsilon_{D^*}^{\lambda *}}{(p_1-p_4)^2-m_{D_{1}}^2+im_{D_{1}}\Gamma_{D_{1}}} \label{d1-b} 
\end{eqnarray}
with  
\begin{eqnarray}
\mathcal{A}^{e^+e^- \to Y(4220)}_{\rho}&=&\bar{v}(k_2)e\gamma_{\mu}u(k_1)\frac{-g^{\mu\nu}em_{Y}^2}{sf_{Y}}\frac{-g_{\nu\rho}+p_{1\nu}p_{1\rho}/m_Y^2}{s-m_Y^2+im_Y\Gamma_Y}, \nonumber
\end{eqnarray}
where $\tilde{g}^{\rho}_{\alpha}=-g^{\rho}_{\alpha}+(p_1^{\rho}-p_4^{\rho})(p_{1\alpha}-p_{4\alpha})/m_{D_{1}}^2$.
Then, the differential cross section of $e^+e^-\to D^{(*)}\bar D^*\pi$ can be expressed by the independent variables of three-body phase space, i.e.,
\begin{eqnarray}
d\sigma=\frac{1}{(2\pi)^5}\frac{\left|\bf{p_2}\right|\left|\bf {p_3}^*\right|\overline{\left|\mathcal{M}^{\textrm{Total}}\right|^2}}{32(k_1\cdot k_2)\sqrt{s}} dm_{D^{(*)}\bar{D}^*}d\Omega_{2}d\Omega_{3}^*
\end{eqnarray}
with
\begin{eqnarray}
\mathcal{M}^{\textrm{Total}}=\mathcal{M}^{Nonpeak}+e^{i\phi}\mathcal{M}^{D_1(2420)}, \nonumber
\end{eqnarray}
where the overline above the total scattering amplitude stands for the average over the spin of electron and positron and the sum over the spin of final states, and the $\bf {p_3}^*$ and $\Omega_{3}^*$ are the three-momentum and solid angle of $\bar{D}^{*0}$ meson in the center of mass frame of the $D^{(*)}\bar{D}^*$ system.

\begin{figure}[ht]
	\includegraphics[width=8cm,keepaspectratio]{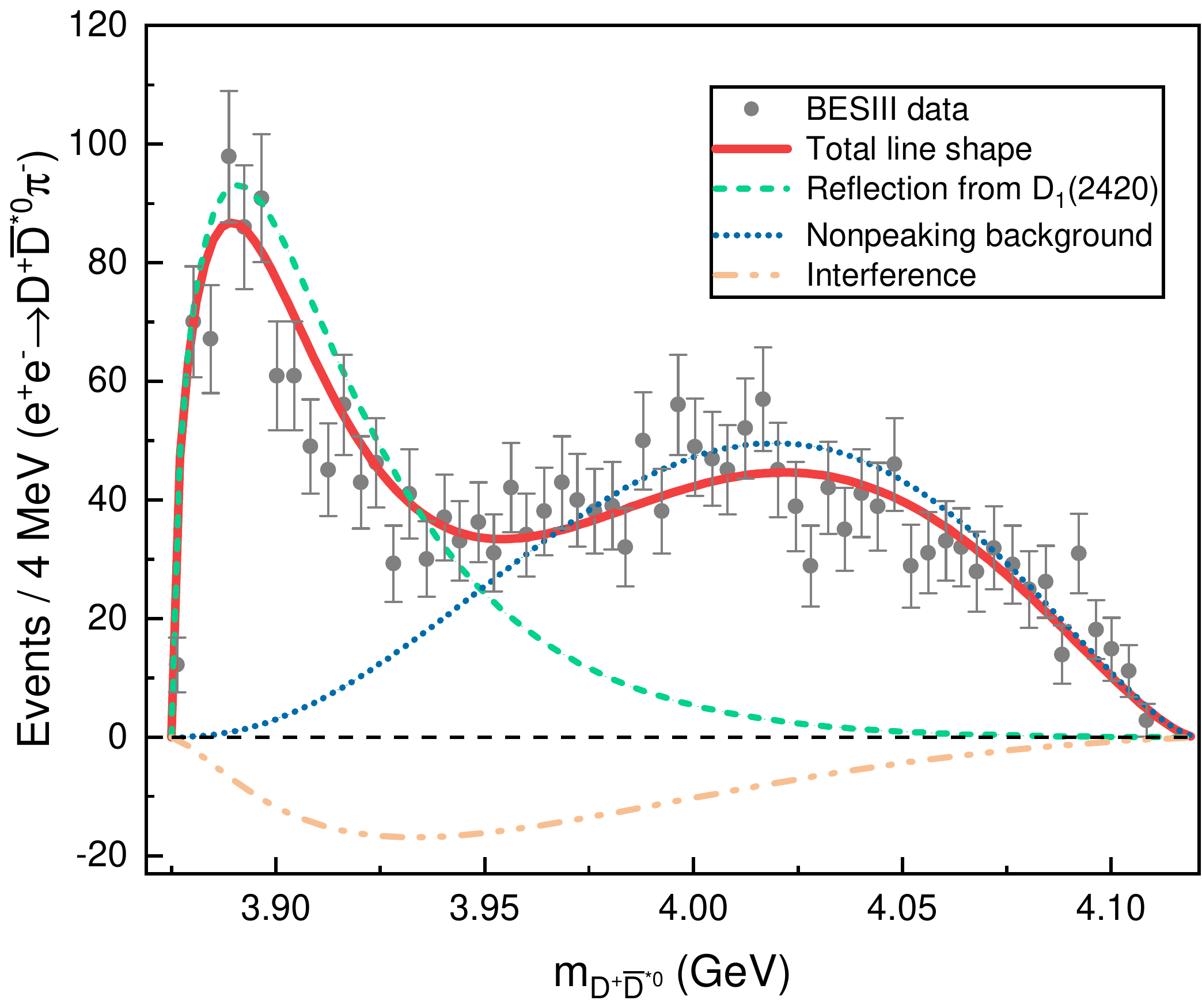}
	\caption{ The fit to the experimental data of $e^+e^-\to D^+\bar D^{*0}\pi^-$ by line shapes on the spectrum of $m_{D^+\bar D^{*0}}$ \cite{Ablikim:2013xfr}, where the structure near the threshold of $D^+\bar D^{*0}$ corresponds to the reported $Z_c(3885)^{+}$. Here, included are only the reflection from charmed meson  $D_1(2420)$ and normal nonpeaking contributions together with their interference. }\label{fig:Line shape1}
\end{figure}

\begin{figure}[ht]
	\includegraphics[width=8cm,keepaspectratio]{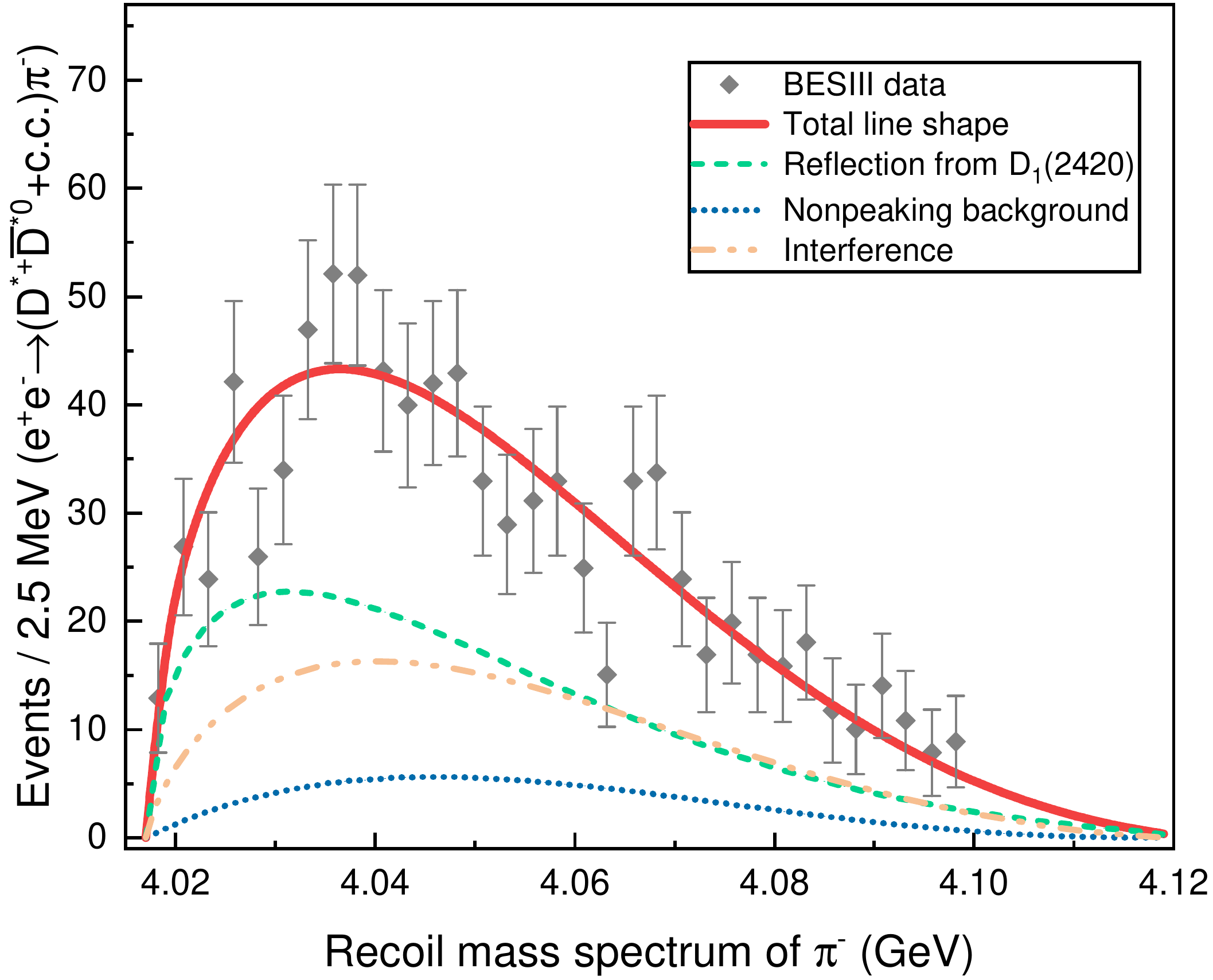}
	\caption{The same as Fig. 2 but for recoil mass spectrum of $\pi^-$ of $e^+e^- \to D^{\ast +} \bar{D}^{\ast 0} \pi^-$. The broad peak corresponds to $Z_c(4025)$. }\label{fig:Line shape2}
\end{figure}

With the above preparations, we can focus on the line shapes of the invariant mass spectra of $m_{D^+\bar D^{*0}}$ and $m_{D^{*+}\bar D^{*0}}$ for the processes $e^+e^-\to D^+\bar D^{*0}\pi^-$ and $e^+e^-\to D^{*+}\bar D^{*0}\pi^-$, respectively, which are directly related to two observed charmoniumlike structures $Z_c(3885)^{\pm}$ and $Z_c(4025)^{\pm}$, respectively. In the present fit, the averaged resonance parameters of $D_1(2420)^-$  in PDG \cite{Tanabashi:2018oca} are taken as input. Here, it is worth emphasizing that besides $D^{(*)}D_1(2420)$ the $S$-wave coupling between $Y(4220)$ and other two-body open charm channels involving a broader charmed meson, such as $D_1(2430)$, may also contribute to the process  $e^+e^-\to D^{(*)+}\bar D^{*0}\pi^-$. However, we have found that they can play a very similar role to the direct contribution of the background. So, for the purpose of reducing the fitting parameters, we refer to the treatment of experimental analysis, i.e., introduction of a factor $(m_{D^{(*)+}\bar D^{*0}}-(m_{D^{(*)+}}+m_{\bar{D}^{*0}}))^a((\sqrt{s}-m_{\pi^-})-m_{D^{(*)+}\bar D^{*0}})^b$ \cite{Ablikim:2013xfr} into nonpeaking amplitudes of Eqs. (\ref{np-a}) and (\ref{np-b}) to absorb their contributions.

{In Fig.~\ref{fig:Line shape1}, we present our fit to the experimental data of $e^+e^- \to D^+ \bar{D}^{\ast0} \pi^-$, where the individual contributions of reflection from $D_1(2420)$, nonpeaking background, and their interference are also descried.  From the individual contributions, one can find that the peak structure near $D^\ast \bar{D}$ threshold dominantly results from the $D_1(2420)$ reflection, while the enhancement in the vicinity of 4.05 GeV comes from the nonpeaking background.  As for $e^+ e^- \to D^{\ast +} \bar{D}^{\ast 0} \pi^-$, our fit to the $D^{\ast + }\bar{D}^{\ast 0} $ invariant mass distributions is presented in Fig.~\ref{fig:Line shape2}. {Here, it is worth mentioning that because the BESIII data here corresponds to the recoil mass spectrum of $\pi^-$, so the contribution from $e^+e^- \to Y(4220) \to \bar{D}^{*0}D_1^0(2420) \to \bar{D}^{*0}D^{*+}\pi^-$ should be included, which will produce an extra factor of 2 in the amplitude of Eq. (\ref{d1-b}). } Our fit indicates that the reflection from $D_1(2420)$ is crucial in the peak structure near the threshold of $D^\ast \bar{D}^\ast$ and the interference between the reflection from $D_1(2420)$ and the nonpeaking background is also non-ignorable.  From the present fit, one can find both $Z_c(3885)$ and $Z_c(4025)$ structures can be reproduced well by including only the reflection of $D_1(2420)$ and normal nonpeaking background together with their interference. Thus, our theoretical analysis indicates that two charmoniumlike structures $Z_c(3885)$ and $Z_c(4025)$ should have the same origin, i.e., the reflection of the $P$-wave charmed meson  $D_1(2420)$. In addition to the perfect reproduction of experimental data by line shapes, we also notice the peak position of the reflection line shape of $D_1(2420)$ in the distribution of invariant mass of $D^+\bar D^{*0}$ is weakly dependent on the input of its resonance parameters. The peak position is almost exactly equal to 3.890 GeV. This may provide us a very natural explanation for the problem of mass inconsistency between $Z_c(3885)^{\pm}$ and $Z_c(3900)^{\pm}$ observed in hidden-charm final states $J/\psi \pi^+$ \cite{Liu:2013dau}, which implies these two $Z_c$ states with close mass may be generated from different production mechanisms. }

\begin{figure}[t]
	\includegraphics[width=8cm,keepaspectratio]{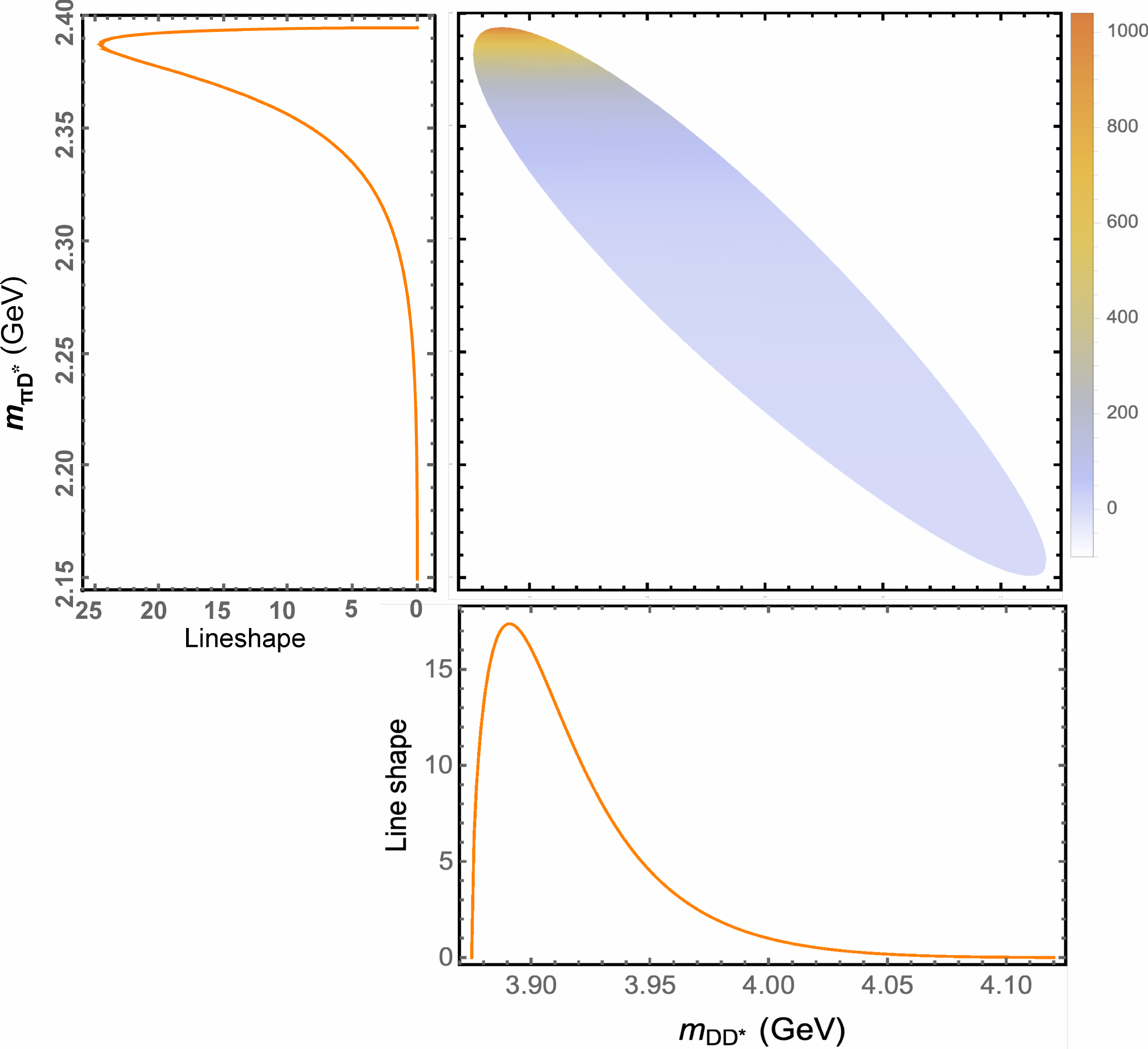}
	\caption{A sketch diagram illustrating the reflection from $D_1(2420)$ in $e^+e^- \to D^+ \bar{D}^{\ast0 } \pi^-$ . }\label{fig:dalitz}
\end{figure}

{From our fit to the experimental data of both $e^+ e^- \to D^{+}\bar{D}^{\ast 0} \pi^-$ and $e^+ e^- \to D^{\ast +} \bar{D}^{\ast 0} \pi^-$, one can find the reflections from $D_1(2420)$ are crucial in revealing the nature of  $Z_c(3885)$ and $Z_c(4025)$. Here, we take $e^+ e^- \to D^+ \bar{D}^{\ast 0} \pi^-$ as an example to further show how the reflection from $D_1(2420)$ affects the recoil mass spectrum of a pion meson. In Fig.~\ref{fig:dalitz}, we show a dalitz plot of $e^+e^- \to D^+ \bar{D}^{\ast0 } \pi^-$ presenting individual contributions from $m_{\pi D^*}$ and $m_{DD^*}$. One can find the signal of $D_1(2420)$ in the dalitz plot, in which the projection of $D_1(2420)$ in the invariant mass spectrum of $D^\ast \pi $ is evident. Moreover, since the signal of $D_1(2420)$ is located at the edge of the phase space, its projection in the $D\bar{D}^\ast$ invariant mass spectrum is concentrated near the threshold of $D\bar{D}^\ast$ as a peak structure. In a similar way, the reflection from $D_1(2420)$ in the process of $e^+e^- \to D^{\ast +} \bar{D}^{\ast 0} \pi^-$ can be analyzed. In Fig. \ref{fig:Line shape1}, we notice that  the reflection peak shape in the invariant mass spectrum of $m_{D^+\bar D^{*0}}$  behaves narrower than that of $m_{D^{*+}\bar D^{*0}}$, which is also consistent with experimental data. The reason for this is due to kinematical properties of the reaction processes, i.e., the threshold of the channel $D^*D_1(2420)$ is more far away from the experimental center of mass energy of 4.26 GeV than $DD_1(2420)$.  }

In fact, in addition to the description for the invariant mass spectrum, the identification of nature of $Z_c(3885)^{\pm}$ can also be achieved by an ingenious method, i.e., the measurement of an angular distribution of a $D$ meson in the center of mass frame of the  $D\bar{D}^*$ final state system. Here, an asymmetric parameter $\mathcal{A}$ reflecting the degree of asymmetry in the angular distribution can be defined as  
\begin{eqnarray}
\mathcal{A}=\frac{\sigma_{\left |\textrm{cos}~\theta_{\pi D}\right |>0.5}-\sigma_{\left |\textrm{cos}~\theta_{\pi D}\right |<0.5}}{\sigma_{\left |\textrm{cos}~\theta_{\pi D}\right |>0.5}+\sigma_{\left |\textrm{cos}~\theta_{\pi D}\right |<0.5}}, \label{eq-asmmetry}
\end{eqnarray}
where $\sigma_{\left |\textrm{cos}~\theta_{\pi D}\right |>0.5}$ and $\sigma_{\left |\textrm{cos}~\theta_{\pi D}\right |<0.5}$ are the integrated cross sections or event numbers of the reaction $e^+e^-\to D^+\bar D^{*0}\pi^-$ in the angle region of $\left |\textrm{cos}~\theta_{\pi D}\right |$ greater than  0.5 and smaller than 0.5, respectively, and $\theta_{\pi D}$ is the angle between bachelor pion and the $D$ meson directions in the rest frame of the $D\bar{D}^*$ system. In general, the mechanism from the direct decay of a charged tetraquark candidate into $D\bar{D}^*$ can present a symmetric angular distribution on $\textrm{cos}~\theta_{\pi D}$, which will lead to $\mathcal{A}=0$. On the other hand, for the contribution from the reflection of charmed meson $D_1(2420)$, the situation will become quite different. In Ref. \cite{Ablikim:2013xfr}, the BESIII measurements show an experimental asymmetric parameter $\mathcal{A}_{\textrm{data}}=0.12\pm0.06$, which is not close to zero. Thus, this data can just provide a test of our non-resonant explanation for $Z_c(3885)$.

\begin{figure*}[htbp]
	\includegraphics[width=15cm,keepaspectratio]{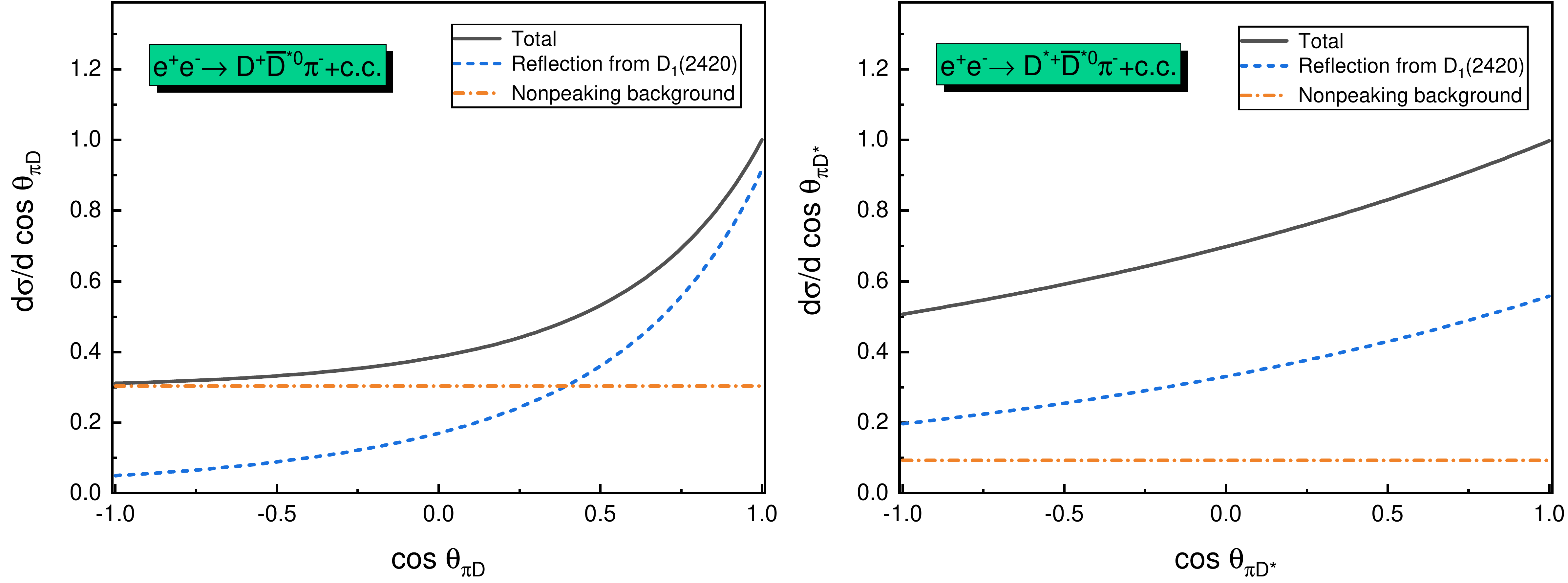}
	\caption{ The predicted differential cross sections of $e^+e^-\to D^+\bar D^{*0}\pi^-$ and $e^+e^-\to D^{*+}\bar D^{*0}\pi^-$ vs. $\textrm{cos}~\theta_{\pi D}$ and $\textrm{cos}~\theta_{\pi D^*}$, respectively. Here, the maximums of angular distributions are normalized to one.    \label{fig:Angulardistribution} }
\end{figure*}

Based on the above fit to experimental data by line shapes, the differential cross sections of $e^+e^-\to D^+\bar D^{*0}\pi^-$ vs. $\textrm{cos}~\theta_{\pi D}$ can be predicted and are shown in Fig. \ref{fig:Angulardistribution}, where the individual contributions from the reflection of $D_1(2420)$ and nonpeaking backgrounds are also plotted. We can see that the angular distribution of a nonpeaking contribution is symmetric, but there appears a monotonically increasing distribution for the reflection of $D_1(2420)$ for the range of $\textrm{cos}~\theta_{\pi D}=-1$ to $\textrm{cos}~\theta_{\pi D}=1$, which corresponds to an asymmetric parameter $\mathcal{A}_{D_1(2420)}=0.269$. After adding up the distributions of reflection of $D_1(2420)$ and nonpeaking background together with their interference term, we find that the total asymmetric parameter can be calculated as
\begin{eqnarray}
\mathcal{A}_{\textrm{Total}}=0.128, \nonumber
\end{eqnarray}
which is almost exactly equal to an experimental value $\mathcal{A}_{\textrm{data}}=0.12\pm0.06$. Therefore, the experimental result of an angular ($\theta_{\pi D}$) distribution of $e^+e^-\to D^+\bar D^{*0}\pi^-$ provides a very strong evidence for our understanding of the charmoniumlike $Z_c(3885)$ structure. 

{We noticed that the BESIII Collaboration once argued that $Z_c(3885)$ cannot be due to the reflection from the $D_(2420)$ \cite{Ablikim:2013xfr}, which is based on a Monte Carlo simulation and the comparison with experimental data of measured asymmetric parameter $\mathcal{A}_{\textrm{data}}=0.12\pm0.06$ \cite{Ablikim:2013xfr}. In fact, as the theorist, we only believe the experimental data provided by the experimentalist. To some extent, exploring the underly mechanism behind the experimental data should be left to the theorist. Thus, we restudied this interesting possibility of $Z_c(3885)$ and $Z_c(4025)$ structures as the reflection from $D_1(2420)$. Obviously, our conclusion is different from that from a simple Monte Carlo simulation \cite{Ablikim:2013xfr}. This difference should be emphasized before making definite conclusion of decoding $Z_c(3885)$ and $Z_c(4025)$. }

We further predict the differential cross section of $e^+e^-\to D^{*+}\bar D^{*0}\pi^-$ vs. $\textrm{cos}~\theta_{\pi D^*}$ in Fig. \ref{fig:Angulardistribution}, where the line shapes of angular distributions for total contribution and reflection of $D_1(2420)$ are very similar to each other, which can correspond to two asymmetric parameters $\mathcal{A}^{\prime}_{D_1(2420)}=0.0334$ and $\mathcal{A}^{\prime}_{\textrm{Total}}=0.0189$ similar to the definition of $\mathcal{A}$ in Eq. (\ref{eq-asmmetry}), respectively. These predictions can be verified in future experimental measurements, which are also helpful for clarifying the nature of the charmoniumlike $Z_c(4025)$ structure.

\begin{table}
  	\caption{ The parameters for fitting the experimental line shape on the invariant mass spectrum of $m_{D^{(*)+}\bar D^{*0}}$ of $e^+e^-\to D^{(*)+}\bar D^{*0}\pi^-$.}
  	\setlength{\tabcolsep}{3.8mm}{
  	\begin{tabular}{cccccc}
			\toprule[1.2pt] 
			Fit &  $e^+e^-\to D^+\bar D^{*0}\pi^-$ & $e^+e^-\to D^{*+}\bar D^{*0}\pi^-$  \\
			\toprule[0.8pt] 
          $\left | \frac{g_{D_{1}{D^{(*)}}Y} g_{{D}_{1}D^*\pi}}{g_{YD^{(*)}{D}^*\pi}} \right |$ & 0.10 $(\textrm{GeV}^{-2})$ & 0.49  $(\textrm{GeV}^{-2})$ \\
            $\phi$ & 5.16  & 3.08   \\
            $a$ & 0.92  & 0.25   \\
            $b$ & 0.52  & 0.98   \\
            $\chi^2/d.o.f$ & 1.76  & 0.99   \\

\bottomrule[1.2pt]

\end{tabular}\label{table:fit}}
  \end{table}

\noindent\textit{Summary.}--In this letter, we have proposed a universal non-resonant explanation to understand the nature of two charged charmoniumlike states $Z_c(3885)^{\pm}$ and $Z_c(4025)^{\pm}$. This novel non-resonant view is completely different from the interpretation by an exotic tetraquark configuration, which has been treated as a mainstream opinion in the previous theoretical discussions \cite{Braaten:2014qka,Maiani:2014aja,Guo:2013sya,Karliner:2015ina,Voloshin:2013dpa,Aceti:2014uea,Wang:2013vex,Deng:2014gqa,Chen:2015ata,Goerke:2016hxf,Patel:2014vua,Wang:2013exa,Qiao:2013dda}. With the help of an effective Lagrangian approach, starting from a specific dynamical reaction $e^+e^-\to D^{(*)}\bar D_1(2420)\to D^{(*)}\bar D^*\pi$, where $ \bar{D}_1(2420)$ is off shell at an experimental energy of $\sqrt{s}=4.26$ GeV, we have found that the intermediate $P$-wave charmed meson $ D_1(2420)$ can produce the line shape of a reflective peak near the threshold of invariant mass spectra both of $m_{D^{+}\bar D^{*0}}$ and $m_{D^{*+}\bar D^{*0}}$. Combined with a reflection of $ D_1(2420)$ and a nonpeaking background, the  experimental signals of $Z_c(3885)^{\pm}$ and $Z_c(4025)^{\pm}$ can be simultaneously reproduced well without introducing any exotic hadron candidates. Furthermore, we have predicted the differential cross sections of $e^+e^-\to D^{(*)+}\bar D^{*0}\pi^-$ vs. $\textrm{cos}~\theta_{\pi D^{(*)}}$, where $\theta_{\pi D^{(*)}}$ is an angle between bachelor pion and the $D^{(*)}$ meson directions in the rest frame of the $D^{(*)}\bar{D}^*$ system. The theoretical results of an angular distribution of $e^+e^-\to D^{+}\bar D^{*0}\pi^-$ show an asymmetric parameter $\mathcal{A}_{\textrm{Total}}=0.128$, which is consistent with experimental value of $\mathcal{A}_{\textrm{data}}=0.12\pm0.06$. This is another compelling evidence to support our novel view for the $Z_c(3885)$ structure as a reflection of the $P$-wave charmed meson $ D_1(2420)$. 

The present studies provide a unique perspective to decode charged $Z_c$ states in the $XYZ$ family, and show again a possible fact of {\it "what you see is not what you get"} in the search for genuine hadron structures. On the other hand, this novel example also reflects the complexity  of resonance phenomena in non-perturbative QCD. Thus, more efforts by theorists and experimentalists are still needed for understanding these $XYZ$ states more deeply.
Of course, we believe that more precise experimental data can eventually answer to these problems, which is worth expecting in the future.

\section*{ACKNOWLEDGEMENTS}

This work is partly supported by the China National Funds for Distinguished Young Scientists under Grant No. 11825503, the National Natural Science Foundation of China under Grant No. 11775050, and by the Fundamental Research Funds for the Central Universities.

\end{document}